\documentclass[10pt,sigconf,letterpaper,nonacm]{acmart}

\usepackage{xcolor}
\usepackage{xspace}
\usepackage{graphicx}
\usepackage{multirow}
\usepackage{subcaption}

\usepackage{lipsum}
\usepackage{enumitem}

\newcommand{\gpts}{GPTs\xspace}
\newcommand{\gpt}{GPT\xspace}

\begin{document}

\title{GPTs Window Shopping: An analysis of the Landscape of Custom ChatGPT Models}

\author{Benjamin Zi Hao Zhao}
\email{ben_zi.zhao@mq.edu.au}
\affiliation{%
  \institution{Macquarie University}
  \city{Sydney}
  \country{Australia}
}
\author{Muhammad Ikram}
\email{muhammad.ikram@mq.edu.au}
\affiliation{%
  \institution{Macquarie University}
  \city{Sydney}
  \country{Australia}
}
\author{Mohamed Ali Kaafar}
\email{dali.kaafar@mq.edu.au}
\affiliation{%
  \institution{Macquarie University}
  \city{Sydney}
  \country{Australia}
}
\renewcommand{\shortauthors}{Zhao et al.}
 
\begin{abstract}
OpenAI's ChatGPT initiated a wave of technical iterations in the space of Large Language Models (LLMs) by demonstrating the capability and disruptive power of LLMs. OpenAI has prompted large organizations to respond with their own advancements and models to push the LLM performance envelope. OpenAI has prompted large organizations to respond with their own advancements and models to push the LLM performance envelope. OpenAI's success in spotlighting AI can be partially attributed to decreased barriers to entry, enabling any individual with an internet-enabled device to interact with LLMs. What was previously relegated to a few researchers and developers with necessary computing resources is now available to all. A desire to customize LLMs to better accommodate individual needs prompted OpenAI's creation of the GPT Store, a central platform where users can create and share custom GPT models. Customization comes in the form of prompt-tuning, analysis of reference resources, browsing, and external API interactions, alongside a promise of revenue sharing for created custom GPTs. In this work, we peer into the window of the GPT Store and measure its impact. Our analysis constitutes a large-scale overview of the store exploring community perception, GPT details, and the GPT authors, in addition to a deep-dive into a 3rd party storefront indexing user-submitted GPTs, exploring if creators seek to monetize their creations in the absence of OpenAI's revenue sharing. 

\end{abstract}

\settopmatter{printfolios=true}
\maketitle

\section{Introduction}
With the rise of machine learning and artificial intelligence, particularly with the emergence of large language models (LLMs), we are witnessing the inception of novel types of services. Much akin to how software developers offset their development expenses by vending their software applications on platforms such as Google Play~\cite{gplay} and Apple's iTunes~\cite{itunes}, customized machine learning model developers follow suit by monetizing or disseminating their developed or fine-tuned machine learning models—referred to simply as models—across various marketplaces such as OpenAI's \gpts store~\cite{OpenAIStore}. In this paper, we analyze the landscape of OpenAI's \gpts shop, arguably the largest platform of user customized LLMs in an accessible, user friendly format. 

The current model for foundational model providers includes the ability to fine-tune their foundational LLM for specific downstream tasks needed by a developer. However, the cost of procuring large quantities of task-specific data and costs of iteratively fine-tuning may still be prohibitive to less resourced developers, hence the emergence of \emph{prompt tuning} as an alternative to guide the output generation of the foundation models. Specifically, through Zero, One, Few-shot learning, and completion-style prompts approaches~\cite{perez2021true,xu2023making,zhao2024lora}, the developer is able to pre-empt a conversation with an LLM with examples of input-response pairs, adjust the tone and format of the output, and provide instructions for the LLM output generation, such as requesting the LLM to think in sequential steps, or based in reasoning. 

The onus for such promoting is on the developer to enact before the end-user begins their conversation or interaction with a foundational LLM or GPT. What distinguishes OpenAI's \gpts platform~\cite{achiam2023gpt} from other foundational model providers such as Google~\cite{team2023gemini}, Mistral~\cite{jiang2023mistral}, IBM~\cite{mishra2024granite}, Meta~\cite{touvron2023llama} and Microsoft~\cite{abdin2024phi} is the integration of these author prompts and resources prior to a conversation with the \gpts directly within the ChatGPT platform. By providing a convenient interface to create custom interfaces, with no requirement for the knowledge of programming or LLM internals, any user with a ChatGPT Plus~\cite{chatgptplus} subscription may create and publish their own custom \gpts, and share it to other ChatGPT plus subscribers. A directory of the most popular \gpts are presented in the \gpts Store~\cite{OpenAIStore}, and a host of other 3rd-party storefronts linking back to the OpenAI platform.

Therefore, the appeal for the OpenAI \gpt store is evident from the perspective of usability. OpenAI's plans to facilitate revenue sharing with \gpt creators, serve as an incentive for creators to produce high-quality \gpts. While this process is still in its infancy, with discussions of monetization originating with the OpenAI's announcement of the platform at DevDay~\cite{introGPTs}. To date, there is still no clear model for revenue sharing, leaving creators eager to expedite monetization have devised creative methods to divert users away from the platform and directly capitalize on their creation. 

In this work we provide analysis into the current landscape of the OpenAI \gpt store, augmented by a view into a 3rd party \gpt store in which \gpts authors have proactively opted-in for inclusion within the directory external to OpenAI. Both analyses serve to illustrate the current engagement of the public (ChatGPT plus subscribers) with the custom \gpts store, what is being published on the \gpt store, to what extent are the \gpt capabilities leveraged in custom \gpts, and with creators seeking to capitalize of their creative efforts, we inspect the authors of these \gpts, their externally linked hosting infrastructure, and how many seek to monetize users off-platform. To foster further research we release our code and data publicly\footnote{\url{https://github.com/gpts-survey/gptsurvey}}.

\section{Dataset}
In this study, we leverage two datasets: the Beetrove dataset~\cite{beetrove2024gpts} as a comprehensive representation of the \gpt ecosystem, and the EpicGPTstore~\cite{EpicStore}, where \gpts have been proactively listed by their creators.
The data sourced in both datasets is in compliance with guidelines outlined by OpenAI and EpicGPTstore's, as specified in their respective \texttt{robots.txt} files\footnote{\url{https://chat.openai.com/robots.txt} and  \url{www.epicgptstore.com/robots.txt}}. Table~\ref{tab:Overalldata} overviews the datasets analyzed in this study. 

\begin{table}[t]
\caption{An overview of the analyzed datasets.}
\label{tab:Overalldata}
\vspace{-3mm}
\centering
\resizebox{0.8\columnwidth}{!}{%
\begin{tabular}{lrr}
\toprule
\textbf{Dataset}          & \textbf{\# of Custom GPTs}      & \textbf{Collection Method} \\
\midrule
Beetrove~\cite{beetrove2024gpts} & 334K & Search Engine Discovery \\
\midrule
EpicGPTstore~\cite{EpicStore}    & 4,186          &  User-submitted index\\
\bottomrule

\end{tabular}
}
\end{table}

\textbf{Beetrove}
The Beetrove dataset~\cite{beetrove2024gpts} was created as a comprehensive repository to foster community collaboration.
%
The dataset curated for the analysis in this work has been performed by web crawling over a series of days, in a manner akin to search engine discovery to locate a total of 349K Custom \gpts. The need for web discovery of custom \gpts is due to a lack of a single index in which all custom GPTs are listed, for example, accessible through pagination, There exists a keyword search for custom \gpts, however with each interaction, only 12 additional related \gpts are displayed.
Following the discovery of these custom \gpts, metadata about each \gpt was crawled by directly visiting the custon \gpt's OpenAI page. As a result, this dataset contains all publicly available information about each \gpts.
We refer the reader to \cite{beetrove2024gpts} for additional detail about the gathering process, and the availability of auxiliary \gpt store data, including periodic monitoring of previously listed \gpts, and daily crawling of the \gpt homepage to track changes in engagement with the most popular custom \gpts.

\textbf{EpicGPTstore}
The EpicGTPStore~\cite{EpicStore} is a 3rd party \gpt store we investigate. This store provides a paginated index of all \gpts submitted to the website. We crawled the entirety of this storefront to obtain the OpenAI links to the listed \gpts. As this storefront does not provide the same granularity of \gpt metadata as found on OpenAI, we further crawled each \gpt's metadata from it is ``About'' tab on OpenAI. This process was completed in early March 2024, yielding 4,186 \gpts, noting 134 of the \gpts on the storefront were inaccessible, or no longer available during subsequent crawling. It is unclear why these GPTs are absent on the OpenAI GPT store, liekly removed by the creators or OpenAI for policy violations.

\section{Overall Analysis}
The focus of this section is the analysis of the overall landscape of \gpts as provided by the Beetrove dataset. We shall first inspect the community perception and engagement with \gpts, followed by specific details of a given \gpt, and finally a look into the authors behind the \gpts.

\subsection{\gpts Community Perception}
In this section we investigate the community's interaction with Custom \gpts as captured through their engagement through conversations, \gpt quality measured through Ratings, and how custom \gpts interact with other languages.

\begin{figure}[t]
  \centering
  \subfloat[GPT Conversations\label{fig:beetrove_conversations}]{\includegraphics[width=0.48\columnwidth]{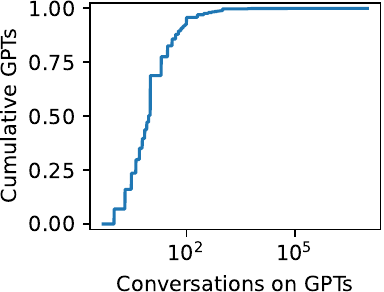}}
  \hfill
  \subfloat[GPT Ratings\label{fig:beetrove_ratings}]{\includegraphics[width=0.48\columnwidth]{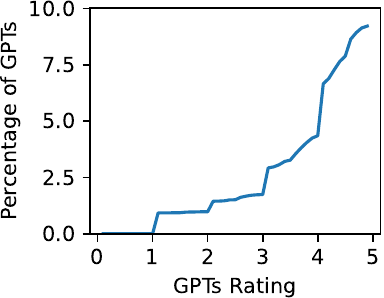}}
  \vspace{-3mm}
  \caption{Cumulative distribution of community engagement with \gpts through the number of conversations with a \gpt and the average rating of \gpts.}
  \label{fig:gpt_engagement}
\end{figure}

\subsubsection{Engagement}
Engagement of the community can be measured by the number of unique conversations had with each of the \gpts. This value is reported for each \gpt, however rounded to retain the most significant orders of magnitude, for example, 323,124, may be rounded to 300K+.
We plot a cumulative distribution of the number of conversations had with each \gpt in Figure~\ref{fig:beetrove_conversations}. From this figure, we observe the overall user engagement with custom \gpts across our dataset. A small 28.5\% of custom \gpts have solicited over 10 conversations.  While on the other extreme, the \gpt store explore page lists the top 12 \gpts of each \gpt category, with consideration of only the 12 of the top \gpts (0.36\% of recorded \gpts), they command approximately, 33.7\% of all custom \gpt conversations recorded within the dataset.

\subsubsection{Ratings}
Another perspective of user perception to the custom \gpts is their rating of said \gpts. For a given \gpt with a rating, it is on average likely to be rated is 4.13. While this may seem to indicate that \gpts are of a high quality, we highlight that the y-axis scale in Figure~\ref{fig:beetrove_ratings} does not conclude at 100\%, as 90.78\% of custom GPTs have not received any ratings. Thus this average rating is only reflective of a small proportion of all \gpts.

\subsubsection{Language}
OpenAI's ChatGPT website is provided in English, with specific optimizations made for the English language~\cite{chen2023phoenix}. While other languages may prompt responses in the same language, the core \gpt 3.5 model from which custom \gpts are based do not claim preserved performance for languages beyond English.
As such, we inspect to what extent do languages other than English feature in custom \gpts. For the analysis of language, we note that there existed examples of \gpts with a combination of languages in both the Description of the \gpt and the Conversation starters. As such we FastText's language detection model on the concatenation of the description and conversation starters to identify the more dominant language.

From Table~\ref{tab:language}, it is clear that English remains the dominant primary language of use, at 86.605\% of all GPTs, the next closest is Spanish at 2.488\%.

\begin{table}[t]
\caption{List of most prominent dominant language of \gpts description and conversation starters.}
\label{tab:language}
\vspace{-3mm}
\centering
\resizebox{0.7\columnwidth}{!}{%
\begin{tabular}{lrr}
\toprule
Language           & GPTs Count      & Percentage (\%) \\
\midrule
English            & 289,562          & 86.605 \\
Spanish; Castilian & 8,317            & 2.488  \\
Japanese           & 7,311            & 2.187  \\
Chinese            & 7,146            & 2.137  \\
French             & 4,579            & 1.370  \\
Portuguese         & 3,678            & 1.100  \\
Korean             & 3,334            & 0.997  \\
German             & 3,133            & 0.937  \\
Others             & 5,112            & 1.539  \\

\midrule
Total             & 334,348            & 100.00  \\
\bottomrule

\end{tabular}
}
\end{table}

\subsection{\gpts Details}

In this section, we analyze the categories in which \gpts are assigned,
Categories are representative of the domain ares in which the \gpt seeks to provide added value. We further investigate the capabilities of \gpts, where capabilities are expanded functionality beyond simple text interaction with the base \gpt model enabled by OpenAI for creators to integrate. We shall finally delve into the utilization of each capability across the Categories.

\subsubsection{Categories}

Categories in OpenAI's \gpts store seeks to catalog and organize the numerous \gpts submitted by users. We remark that multiple attempts to categorize \gpts within the community have spawned third party stores, and listing sites with their own bespoke system of labeling these \gpts. While these alternative, and often more granular lists of categories exist, for our analysis, we use only the official categories provided by the OpenAI store. These categories include: Image Generation (DALLE), Education, Lifestyle, Productivity, Programming, Research, Writing, Other, and the ability to not specify a category.

From Table~\ref{tab:categories}, we can observe nearly half (45.62\%) of \gpts do not have a reported category. With a low 1.39\% being classified specifically for DALLE image generation. A similarly low proportion at 2.48\% were classified for programming, despite the large uptake in the use of AI to assist programming workloads~\cite{becker2023programming}, we posit this may be a consequence of greater emphasis on users leveraging task-specific foundation models fine-tuned specifically for programming tasks~\cite{feng2020codebert,barke2023grounded}, due to their tight integration within existing code editors (e.g. GitHub Copilot). We remark there exists empirical evidence demonstrating that prompt tuning's potential to surpass the performance of fine tuning for programming tasks~\cite{wang2022no}.

\begin{table}[t]
\caption{List of \gpts store categories of user submitted \gpts, the average user rating (excluding unrated \gpts) of each category is provided.}
\label{tab:categories}
\vspace{-3mm}
\centering
\resizebox{0.66\columnwidth}{!}{%
\begin{tabular}{lrrc}
\toprule
Category     & Count  & Percent (\%) & Avg. Rating\\ \midrule
None         & 152,539 & 45.62 & 4.10 \\
Other        & 88,276  & 26.40 & 4.15 \\
Lifestyle    & 20,474  & 6.124 & 4.26 \\
Education    & 19,572  & 5.854 & 4.30 \\
Productivity & 18,588  & 5.559 & 4.11 \\
Research     & 12,159  & 3.637 & 4.07 \\
Writing      & 9,813   & 2.935 & 4.25 \\
Programming  & 8,288   & 2.48 & 4.08 \\
DALL·E       & 4,639   & 1.39 & 3.77 \\ \bottomrule
\end{tabular}
}
\end{table}

\subsubsection{Capabilities}
There are 5 Capabilities requested and leveraged by a custom \gpts  presented in Table~\ref{tab:capabilities}, these capabilities include 1) Browsing, access to the internet, 2) DALLE Images, image generation through OpenAI's DALLE network, 3) Data Analysis, in the form of executing python code, 4) Actions, through APIs to interact with externally hosted resources, and finally 5) Native, using only the \gpt model with an absence of any additional capabilities. With the exception of Native, multiple capabilities may be selected simultaneously.

A significant majority of custom GPTs have enabled the Browsing capability, with DALLE image generation present in a smaller, albeit still significant proportion of \gpts. The least used capability is Actions, at only 2.96\%, we speculate this is a result of the high technical and resource burden required to create and host external functions made accessible to the custom \gpt through an API. Despite this low overall update of Actions, 7 of the top 12 most conversed \gpts use the Action capability, presenting a potentially causal relationship behind the increased number of conversations, and the augmentation of the custom \gpt through external functionality.

We further breakdown the use of each Capability by their use in each different category. Table~\ref{tab:cap_cat} provides this view. From the 334K \gpts,

\begin{table}[t]
\caption{List of capabilities available to a \gpt creator. The Data Analysis capability permits the execution of python code, while Actions provides the capability to integrate external APIs.}
\label{tab:capabilities}
\vspace{-3mm}
\centering
\resizebox{0.66\columnwidth}{!}{%
\begin{tabular}{lrrc}
\toprule
Capability    & Count  & Percent (\%) & Avg. Rating \\ \midrule
Browsing      & 312,898 & 93.58 & 4.13 \\
DALL·E Images & 291,188 & 87.09 & 4.13 \\
Data Analysis & 165,339 & 49.45 & 4.12 \\
Native        & 10,737  & 3.21 & 4.16 \\
Actions       & 9,891   & 2.96 & 4.02 \\ \bottomrule
\end{tabular}
}
\end{table}

\begin{table}[t]
\caption{The proportion of different categorized \gpts, and their utilization of capabilities within OpenAI.}
\label{tab:cap_cat}
\vspace{-3mm}
\centering
\resizebox{1\columnwidth}{!}{%
\begin{tabular}{l|rrrrr|r}
\toprule
             & Browsing & DALL·E & Analysis & Native & Actions & Totals\\ 
             \midrule
None         & 143,849   & 135,386        & 94,867         & 4,516   & 2,468    & 152,539\\
Other        & 82,717    & 77,124         & 32,542         & 2,919   & 3,420    & 88,276 \\
Lifestyle    & 18,897    & 17,891         & 6,078          & 573    & 965   & 20,274  \\
Education    & 18,286    & 16,662         & 7,718          & 745    & 459   & 19,572  \\
Productivity & 17,143    & 15,476         & 7,775          & 841    & 1,017  & 18,588  \\
Research     & 11,396    & 9,772          & 5,942          & 314    & 868   & 12,159  \\
Writing      & 8,968     & 8,241          & 3,252          & 583    & 261   & 9,813  \\
Programming  & 7,772     & 6,092          & 5,744          & 211    & 324   & 8,288  \\
DALL·E        & 3,870     & 4,544          & 1,421          & 35     & 109  & 4,639   \\ \midrule
Totals & 312,898 & 291,188 & 165,339 & 10,737 & 9,891 & 334,348 \\ \bottomrule
\end{tabular}
}
\end{table}

\begin{figure*}[t]
  \centering
  \includegraphics[width=0.95\textwidth]{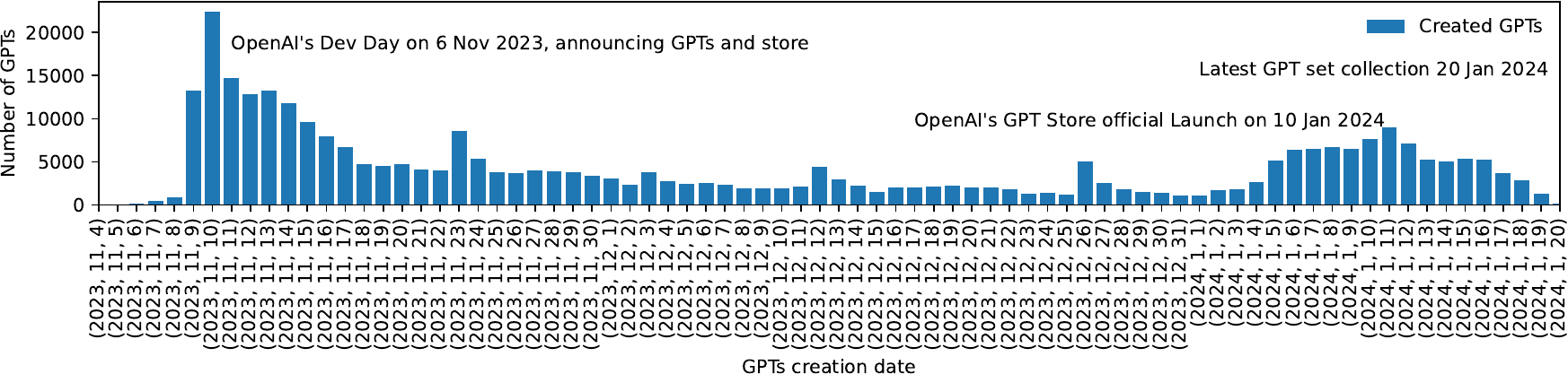}
  \vspace{-3mm}
  \caption{Distribution of GPTs creation dates. Notable events marked as 6 Nov 2023, and 10 Jan 2024 for OpenAI's Dev Day and the Official \gpts store launch respectively. Newest \gpts collected on 20 Jan 2024.}
  \label{fig:beetrove_creation_date}
\end{figure*}

\begin{figure}[t]
  \centering
  \includegraphics[width=0.85\linewidth]{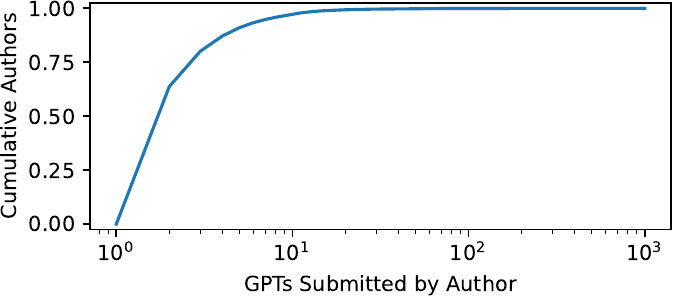}
  \vspace{-3mm}
  \caption{Cumulative distribution of \gpts submitted by the same author. Observe the long tail with a few authors submitting substantial numbers of GPTs.}
  \label{fig:beetrove_author_gpts}
\end{figure}

\subsubsection{Creation}

We plot the distribution of when the \gpts were first created on the \gpts store.
From Figure~\ref{fig:beetrove_creation_date} we can observe a peak of \gpts created following 6 Nov 2023, OpenAI's 2023 DevDay, on which GPTs were announced, and enabled for ChatGPT Plus subscribers to create and share with other subscribers~\cite{introGPTs}. There exists an increase in the base number of \gpts created in the lead up to the official launch of the \gpts store on 10 Jan 2024. This demonstrates that a large degree of GPT creation activity has been spurred on by large event, publicizing and encouraging the active use of OpenAI's custom \gpt service.

\subsection{\gpts Authorship}

\subsubsection{Author Contributions}

We provide in Figure~\ref{fig:beetrove_author_gpts} a view of how there exists a comprehensive spread of \gpt authors, from individual users testing the custom \gpt interface, to large scale authors publishing orders of magnitudes more \gpts, with 10 Authors out of 131K Authors submitting more than 1000 \gpts.

\subsubsection{Author External Links}
In this paragraph we provide metrics on the the external links provided to OpenAI, which are subsequently advertised next to the author information of each submitted custom \gpt. External links can be viewed as a mechanism for \gpt authors to pull users away from the OpenAI platform to engage with the author's other endeavors, for example a demonstration of their prompt engineering for job-seeking, activity on related public repositories, or engagement with their external site, which as we shall investigate in Section~\ref{sec:monetization}, seek to externally monetize the custom \gpts.

\section{3rd Party Store Author Case Study}

In this section we take a closer inspection for a 3rd party \gpt store, \url{https://www.epicgptstore.com/}. While this website is unaffiliated with OpenAI, the \gpts on this storefront are notably different from those generally present on OpenAI (captured by the previous analysis), as the EpicGPTstore requires proactive action by \gpt creators to submit and list their own \gpts on the storefront. As such the set of \gpts this dataset represents contains \gpt author who may be additionally motivated to see their \gpt gain additional community engagement.  As such, our analysis will inspect the types of authors present in the 3rd party \gpt store, before diving into domain based authors, and if there exists attempts by these \gpt authors to monetize their custom \gpts through their external domains in the absence of direct revenue sharing from OpenAI.

\subsection{Author Types}
In this section, we characterize the types of authors observed in the EpicGPTstore marketplace.
We observe 4 notable types of authors, namely, Names, Emails, X (formerly Twitter) handles, and Domains.
Broadly speaking, both X handles and Domains allow a user to quickly explore expanded offerings from the \gpt author.
In total, we obtained 4186 \gpts, authored by 1863 unique authors with 1412 containing Names or other text, 439 Domains and 12 X handles.

\subsubsection{Domain Authors}
\label{subsec:domains}
With these 439 domains, we obtain valid \texttt{whois} data from 383. What is interesting to observe is if domains were specifically registered as a response to capitalize on OpenAI's release of custom \gpts. Notably, the creation dates have been removed from \texttt{whois} information for 20 ``.ai'' Top-Level Domains (TLD).

\begin{figure}[t]
  \centering
  \includegraphics[width=1\linewidth]{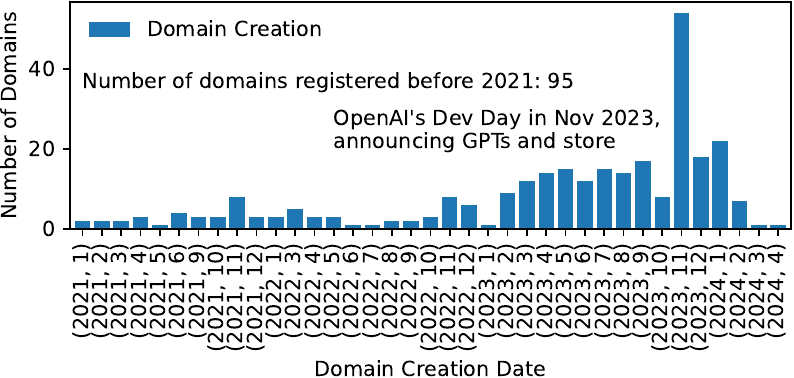}
  \vspace{-3mm}
  \caption{Distribution of domain \gpt authors and their creation date as recorded by \texttt{whois} information. November 2022 is the point in which ChatGPT was first launched by OpenAI.}
  \label{fig:domain_creation_dates}
\end{figure}

\textbf{Creation Date.}
In Figure Figure~\ref{fig:domain_creation_dates}, we observe an elevated trend of registered domains since ChatGPT was released for public use in November 2022, indicating the existence of individuals or organizations registering new domains to position themselves to create new ventures and capitalize on the rise in public awareness of generative AI, and it's capabilities. This increasing trend spikes during November 2023, during the month the ChatGPT store was announced by OpenAI, to enable many more users to create \gpts through an intuitive interface. Before this, creators seeking to create customized \gpts would have relied solely on OpenAI's API (we remark that such early custom model developers may have also privately hosted models of differing architectures and/or with different fine-tuning data independent of OpenAI infrastructure).

\textbf{TLD.}
An analysis of the Top-Level Domains (TLD) of these authors indicates 49.0\% (213) with `.com', 10.3\% (45) with `.ai', and 4.6\% (20) with `.io' as the top 3 TLDs. The presence of the `.ai' TLD indicates a strong focus on capitalizing on the artificial intelligence momentum, while the dominant `.com' domains may be existing online organizations seeking to augment their existing processes with these custom \gpts, we present the lesser used TLDs in Figure~\ref{fig:top_domains} of the Appendix~\ref{appendix:domains}.

\textbf{Hosting Infrastructure}
The hosting infrastructure used by these domains can be inspected to determine if these sites are provided on standard cloud services, or if the domain is hosted on their infrastructure. This may indicate the scale of their operations and the extent of delivering GPT-related tools. To this end, we use the Pythia framework~\cite{Matic_2019} to retrieve the IPs associated with GPTs creators' domains. Pythia utilizes \textit{whois} services and the Registration Data Access Protocol (RDP) to ascertain domain ownership details. Subsequently, we enhance our dataset by integrating metadata such as AS numbers (ASNs) and geolocation information. To accurately map AS numbers, we utilize the BGP Route Views dataset~\cite{routeview}. For country mapping, we rely on the MaxMind~\cite{maxmind} and Potaroo~\cite{potaro} datasets, providing country codes for IP addresses.

Our analysis shows an overwhelming majority of these domains are hosted on infrastructure based in the US (79.1\%), this is reinforced by the presence of large cloud providers in Table~\ref{tab:hosted_domains}. Curiously Japan is the second largest host of these domains (4.0\%), while not in the Top 10, the 16 domains are spread across GMO Internet and Xserver, two Japanese hosting providers. We provide a larger histogram of countries in Figure~\ref{fig:domain_country}, in the Appendix~\ref{appendix:domains}.

\begin{table}[t]
\caption{Top 10 Autonomous System (AS) organizations (Internet Service Providers (ISPs)), and the corresponding number of hosted Domains (\#D).}
\label{tab:hosted_domains}
\vspace{-3mm}
\tabcolsep=0.05cm
\centering
\resizebox{0.75\columnwidth}{!}{%
\begin{tabular}{lrlr}
\toprule
\textbf{AS Organization (ISP)} & \#D & AS Org. (ISP) & \#D \\ \midrule
AMAZON-02  & 114    & NAMECHEAP-NET  &  13        \\
CLOUDFLARENET &  83  & SQUARESPACE  & 13        \\
AMAZON-AES & 22     & AS-HOSTINGER  &  12         \\
GOOGLE-CLOUD-PLATFORM  & 14  & FASTLY & 11        \\
WIX\_COM & 14       & GOOGLE & 10              \\\bottomrule
\end{tabular}
}
\end{table}

\textbf{VirusTotal Analysis}
In this Section, we obtain reports about a given domain author from VirusTotal.
VirusTotal is an online service that provides scanning services for software binaries, and web resources. Their analytics include the aggregation of popular Anti-Virus (AV) scanning tools, with software decompilation locating command and control network infrastructure, and if domains/IPs have been observed engaging with such malicious enterprises.

Through the VirusTotal API, a report was obtained for each domain. For our analysis, only 439 domains returned a virus total report, with a majority of the domains (not being flagged as malicious or suspicious by any scanner. 89 (20.3\%) of the scanned domains are reported as either malicious or suspicious by at least one AV. In Table~\ref{tab:vt}, we report the domains that are reported as malicious by two different AV scanners. This errs on the lower side with the consideration that at the time of writing 92 scanners were evaluated.

\begin{table}[t]
\caption{VirusTotal scans flagging domains as Malicious. A total of 92 scanners were in use for VirusTotal at the time of writing. \# Malicious refers to the number of AV scanners flagging a domain as Malicious.}
\label{tab:vt}
\vspace{-3mm}
\centering
\resizebox{1\columnwidth}{!}{%
\begin{tabular}{lrrrr}
\toprule
\textbf{Domain}	&	\textbf{\# Malicious} & \textbf{Category} & \textbf{ISP} & \textbf{CC}\\
\midrule
gptjp.net	&	6	& Other & CLOUDFLARENET & US\\
citibankdemobusiness.dev	&	3 & Other & AMAZON-02 & US	\\
gantrol.com	&	2 & Education & CLOUDFLARENET & US	\\
usevisuals.com	&	2	& Writing & AMAZON-02 & US\\
engineer.vision	&	2	& Res. \& Analysis & AMAZON-AES & US\\
bahouprompts.com	&	2	& Writing & NAMECHEAP-NET & US\\
gptmakerspace.com	&	2 & Other & AMAZON-AES & US	\\
gptplugins.xyz	&	2 & Education & CLOUDFLARENET	& US\\
\bottomrule
\end{tabular}
}
\end{table}

\subsubsection{External Monetization}
\label{sec:monetization}

To establish is monetization is prevalent in these external domain authors, we visited each of their sites and searched for mentions of \{pricing, price, \$, trial, service, `try it free', 'try it for free', 'start for free', fee, cost, subscription, product\} to determine if the site was seeking to direct users towards paid products or services. We do note that the presence of these terms does not explicitly imply OpenAI services were on-sold to external users, as we do not engage with these providers to gauge their paid offerings. As a result of earlier analysis in the dominance of English \gpts, we do not translate non-English webpages, and the numbers reported may be lower than the true value.

\begin{figure}[t]
  \centering
  \includegraphics[width=1\linewidth]{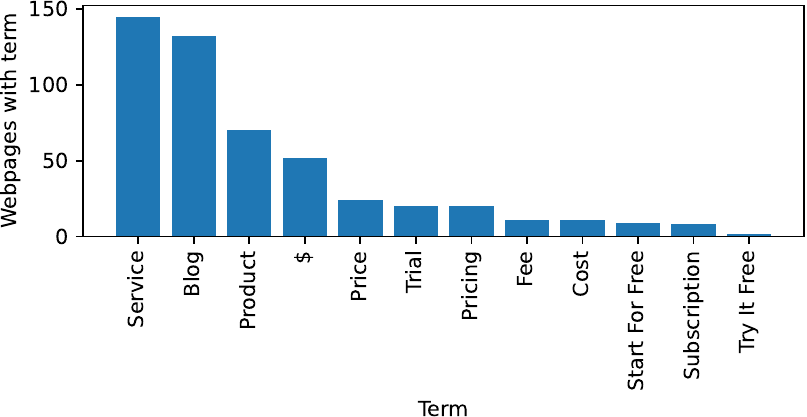}
  \vspace{-3mm}
  \caption{Histogram of terms related to monetization, or ``Blog'' present within the webpages of our \gpt domain authors.}
  \label{fig:money_terms}
  \vspace{-1cm}
\end{figure}

From our 439 domains, 32 could not be resolved, 5 returned a 403 Forbidden response, and 6 were parked domains. Of the remaining 396, 193 (48.7\%) contained at least one of the aforementioned terms related to monetization.
Further, 132 (33.3\%) of these valid sites contained a mention of a `blog', as a means to promote the author's other content. We provide a breakdown of the present terms in Figure~\ref{fig:money_terms}.

\section{Related works}

The most related work to our own is \cite{zhang2024first}, with an early attempt at measurements of a dataset of \gpts crawled from OpenAI, and \gpts on a 3rd party store. An additional angle provided by Zhang et al. is the premise of inferring the initial starting prompt used to prompt-tune a custom \gpt model. As these prompts require careful conscious crafting to maximize the performance of the tuned \gpt, these initial prompts is where the value of the custom \gpt lies. By inferring an initial prompt, an adversary may re-host the \gpts, thereby redirecting any potential traffic or revenue away from the original creator. In this work, we take a deeper dive into a larger dataset of OpenAI \gpts, while providing a closer inspection of the web infrastructure and monetization tactics behind named website creators.

A position paper has furnished a comprehensive summary of the information provided by \gpts and how actions \gpt providers should consider in their threat landscape~\cite{zhao2024llm}. Zhao et al. present real security and privacy threats that LLM providers will need to respond to, however, such actions would safeguard any LLM on the platform, without any consideration for the specific tasks or customization applied to a base model. Of specific relevance to custom \gpts and as demonstrated by \cite{zhang2024first}, prompt-tuned \gpts may leak to a user the initial customized prompt, whereby imitators could simply copy the leaked prompt and re-create the custom \gpt diverting traffic (and potentially revenue) from the original.

Surveys of stores or marketplaces are not new, for example, measurement studies exist for both the Google Play Store~\cite{viennot2014measurement}, and Apple's iOS store~\cite{ali2017same}. While these stores have a similar rating system, for applications (out of five stars) both Google and Apple's application store permit users to leave written reviews of applications, which themselves have been the subject of deeper analysis~\cite{genc2017systematic,fu2013people}.

\section{Conclusion}
In this work, we have analyzed the current landscape of the OpenAI custom \gpt store through a large representative collection of 334K \gpts, finding an overwhelming dominance of English \gpts, an underutilization of OpenAI's categorization system, and a strong correlation of \gpt creation coinciding with OpenAI's two major public accouncements in relation to the \gpt Store. We have further investigated the interactions of \gpt authors who have proactively promoted their \gpts on the 3rd-party EpicGPTstore to reveal a substantial proportion of authors seeking to drive traffic to blogs, and externally monetized web services.

\bibliographystyle{ACM-Reference-Format}
\bibliography{refs}

\appendix

\section{Ethics}

This work investigates user-created custom \gpts submitted to the OpenAI ChatGPT platform and the EpicGPTStore. While human creators have intentionally provided personal details such as their names and/or linked webpages, we describe the bulk of users through statistics, while minimizing circumstances where specific examples are singled out. Our crawling to obtain this data respects the respective \texttt{robots.txt} of each website.

\begin{figure}[t]
  \centering
  \includegraphics[width=0.9\linewidth]{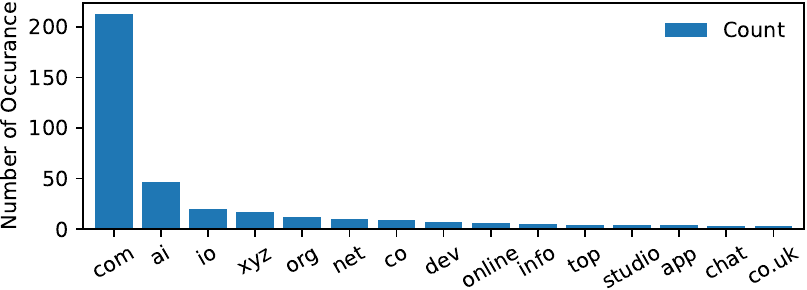}
  \caption{Histogram of Top-Level Domains for listed domain authors (cf. \S~\ref{subsec:domains}).}
  \label{fig:top_domains}
\end{figure}

\begin{figure}[t]
  \centering
  \includegraphics[width=1\linewidth]{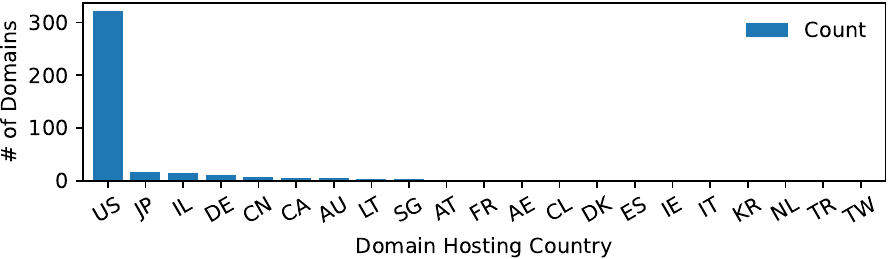}
  \caption{Histogram of country location of hosting infrastructure for domain authors (cf. \S~\ref{subsec:domains}).}
  \label{fig:domain_country}
\end{figure}

\section{Top Level Domains (TLD) of domain \gpt authors}
\label{appendix:domains}
In Section~\ref{subsec:domains}, we analyze the domains of the custom GPTs in our dataset. In Figure~\ref{fig:top_domains}, we provide a larger perspective of TLDs found in the domain \gpt authors.

\section{Country Location of hosting infrastructure for domain authors.}
Figure~\ref{fig:domain_country} provides a larger perspective of countries found in the hosting infrastructure of the domain authors of \gpts. Overwhelmingly, the US has the largest presence, owing to the number of public web hosting providers in the US.

\section{Monetization Terms}
\begin{figure*}[thb]
  \centering
  \includegraphics[width=0.9\linewidth]{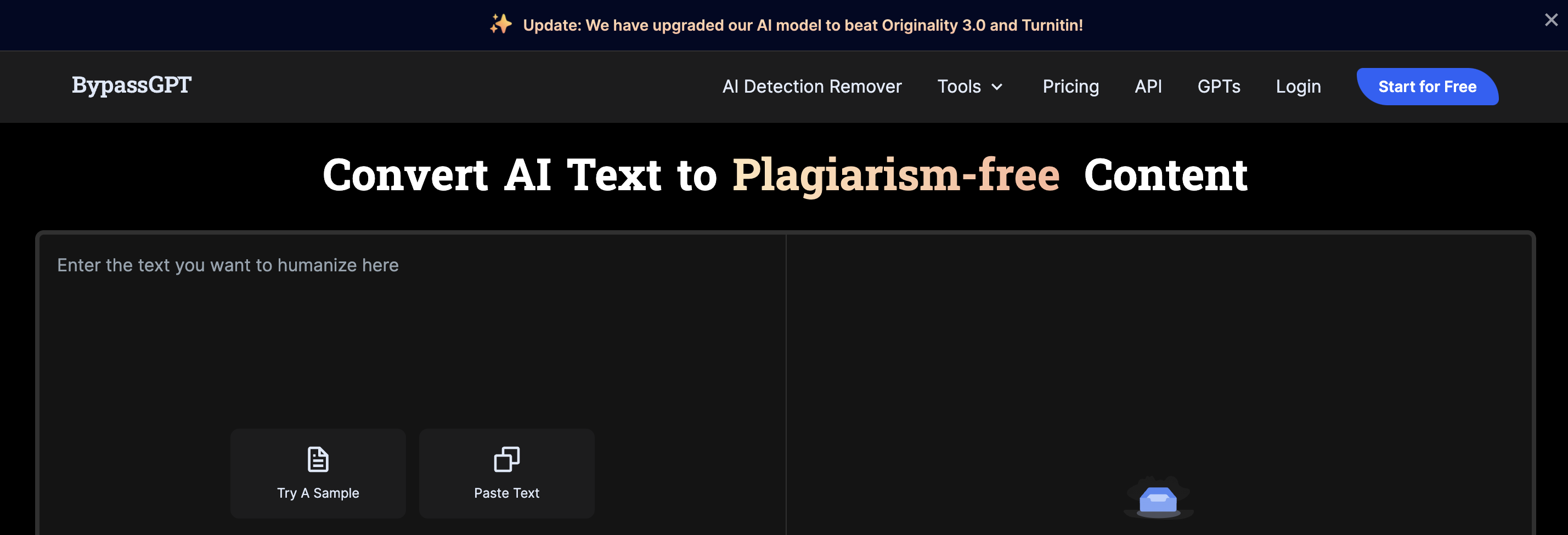}
  \caption{The bypassgpt.ai webpage, with ``pricing'' in the navigation bar.}
  \label{fig:bypassgpt-ai}
\end{figure*}

\begin{figure*}[thb]
  \centering
  \includegraphics[width=0.9\linewidth]{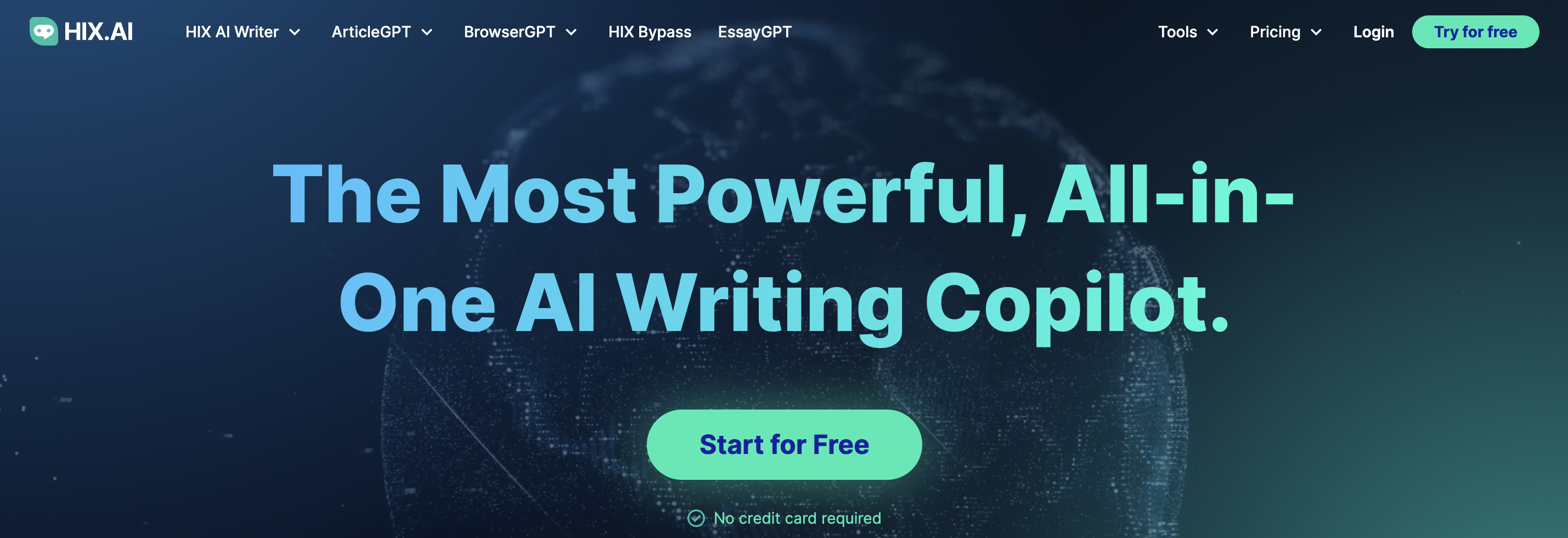}
  \caption{The hix.ai webpage, with ``pricing'' and ``try for free'' in the navigation bar.}
  \label{fig:hix-ai}
\end{figure*}

\begin{figure*}[thb]
  \centering
  \includegraphics[width=0.9\linewidth]{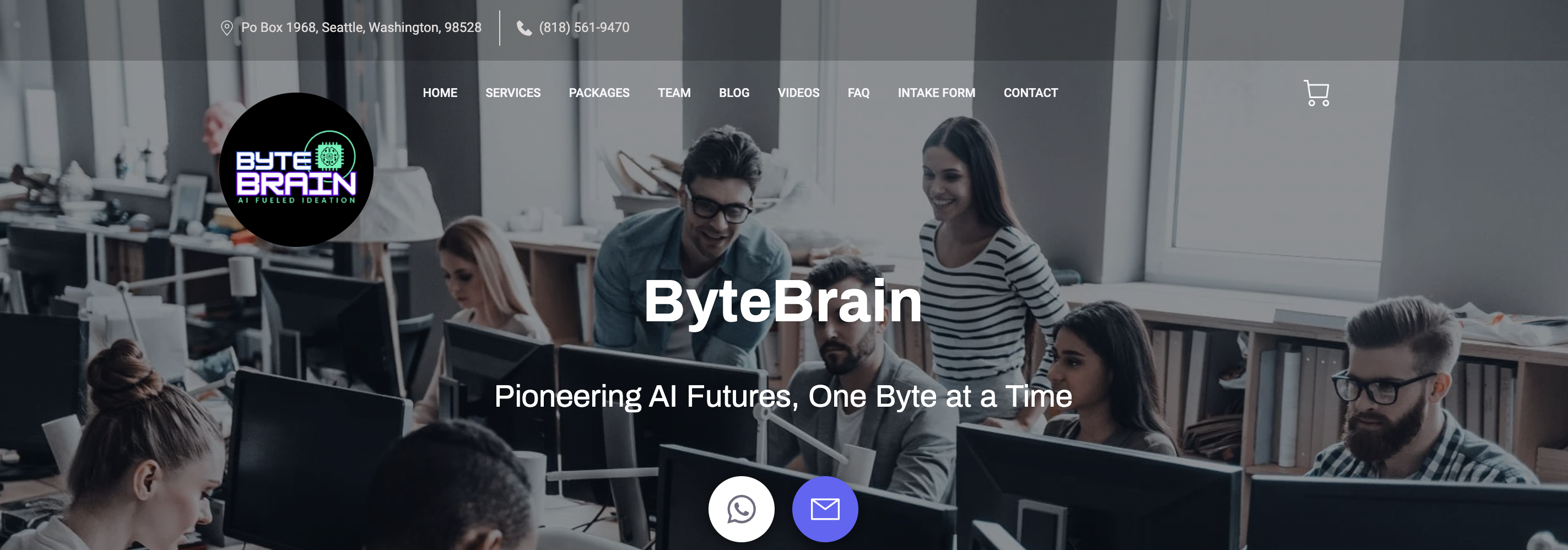}
  \caption{The bytebrain.org webpage, with ``services'' and ``blog'' in the navigation bar.}
  \label{fig:bytebrain-org}
\end{figure*}

In this appendix, we provide visual examples of webpages in Figure~
\ref{fig:bypassgpt-ai}, \ref{fig:hix-ai}, \ref{fig:bytebrain-org}, that we encountered that provided motivation for the list of monetization terms used previously in Section~\ref{sec:monetization}.

\end{document}